\documentclass[amsmath,amssymb,aps,prapplied,twocolumn,superscriptaddress]{revtex4-2}

\usepackage[hyperindex,breaklinks]{hyperref}
\usepackage{graphicx}
\usepackage{graphicx}
\usepackage{dcolumn}
\usepackage{bm}

\usepackage{amsfonts}
\usepackage{amsbsy}
\usepackage{savesym}
\usepackage{latexsym}

\savesymbol{captionbox}
\restoresymbol{SUB}{captionbox}
\usepackage{floatrow}
\usepackage{placeins}
\usepackage{makebox}
\usepackage{microtype}
\begin{document}

\title{Afterpulsing Effect on the Baseline System Error Rate and on the Decoy-State Quantum Key Distribution Protocols
}


\author{Christos Papapanos, Dimitris Zavitsanos,  
	Giannis Giannoulis, Adam Raptakis, Christos Kouloumentas, Hercules Avramopoulos
}

\affiliation{Photonics Communication Research Laboratory,
	School of Electrical and Computer Engineering, National Technical University of Athens, Iroon Polytechniou Street 15780 Greece}

\begin{abstract}
There is considerable interest in predicting the efficiency of Quantum Key Distribution (QKD) protocols when one of their implementation quantities is modified. One significant imperfection that affects the efficiency of the setup is the afterpulse phenomenon which consists in the spontaneous detections triggered by trapped carriers after previous avalanches at the detectors. While it is widely studied in bibliography for various QKD protocols, it has been reported much more scarcely for the well-known decoy-state QKD protocols and for dual detectors only. We develop a theoretical analysis of afterpulsing effect on the decoy-state QKD protocols for multiple detectors, delivering results which can be used as a guide for every practical decoy-state QKD protocol implementation in real-world deployments. A new formula connecting the baseline system error rate and the afterpulse probability is derived which may hold for all protocols as it is consisted of only setup-related quantities. Numerical simulations addressing the significance of breaking down the quantities pertaining to the decoy-state QKD protocols are being made, focusing on the weak+vacuum decoy-state QKD protocol as a characteristic subcase.
\end{abstract}


\maketitle

\section{Introduction}
\label{sec:intro}
The decoy-state Quantum Key Distribution (QKD) protocol \cite{Decoy} is being widely used from fiber-optic communications (ex. SECOQC project) to satellite communications \cite{Liao_2017}. In the field of practical QKD system deployments, where highly attenuated laser sources are used to provide the single photons on the Alice stations, the decoy-state QKD implementation has been proven as one of the most resistant protocols against the photon-number-splitting (PNS) attack \cite{Hwang}, which is linked with the non-zero probability of emitting multi-photon states.  Even though the use of superconducting single photon detectors (SSPDs) has been proposed as an alternative path to achieve long distance QKD with decoy states \cite{Rosenberg}, these expensive detectors based on complex and bulky cryogenic systems cannot be a practical solution offering massive-scale adoption in future QKD deployments. Whilst commercially available SSPD units exhibit no afterpulsing and high detection efficiencies combined with low dark count rates \cite{link1}, most practical and low-cost implementations will probably continue to make use of single-photon avalanche detectors (SPADs). 

Whilst the advances in SPAD technology allow for high detection efficiencies and robust operation at even lower prices \cite{Zhang}, the afterpulse phenomenon or afterpulsing limits the performance of its use. Afterpulsing is the probabilistic phenomenon of triggering a new detection after a previous detection has occured. Afterpulse events do not differ from actual signal-induced events and thus they cannot be distinguished from the latter ones \cite{Juan}. Although there are various models for describing the statistics of afterpulsing \cite{Ferreira_da_Silva_2011,article2}, they all conclude in capturing it via one single quantity; the probability ($p_{\rm AP}$) of having an afterpulse event conditioned on a previous detection event as indicated in ~\cite{afterpulse}. Those results are not correlated with the kind of the protocol or the setup that will be used.

The contribution of afterpulsing to error counts has been addressed, especially in demonstrations where QKD systems operating at GHz detection rates \cite{Gigahertz}. Aside from these high-speed QKD demonstrations, the need to study the afterpulsing effects not only on the Secure Key Rate (SKR) of each protocol but also on every underline quantity is still important for maintaining a wider sense of things; this approach will permit to obtain an overview over the physical meaning behind every implementation quantity of the QKD protocol. In this direction, significant work has been reported in other papers for other protocols such as BB84 QKD but their method cannot be used in decoy-state protocols, see \cite{eraerds2010quantum,Afterpulse_other_protocol}. 

From the deployment perspective of decoy-state QKD implementations, it is important to formulate the way that the afterpulse phenomenon affects the efficiency of the QKD protocols based on closed-form expressions. A great work for the case of multiple protocols- applied also to decoy-state protocols- is the work of Fan-Yuan et al. in \cite{Yuan} where they have focused on a dual detector setup at Bob's side (the receiver). In our work, we aim to generalize the Bob's setup, as the model considered in our paper does not set limitations for the setup's number of detectors. Nevertheless, our analysis is restricted only to decoy-state protocols instead of multiple protocols as in \cite{Yuan}. Aside from the compatibility of our proposed model with different receiver layouts, we also focus on the effect of afterpulsing to the baseline system error rate. More specifically, we derive a mathematical expression between the afterpulse probability and the baseline system error rate, where tradeoff-like relation between the dark count rate and the afterpulse probability for generating the same Quantum Bit Error Rate (QBER) is revealed. The mathematical form of this relation is important to know as the QBER establishes an upper limit for quantum error correction algorithms above which these algorithms cannot be used.

Compared to the theoretical outcomes in \cite{Yuan}, as proved in \cite{Dual1} and also studied in \cite{Dual2}, quiet (low noise), slow detectors exhibit similar or better performance than dual detectors at longer distances (longer than $100$ km), while noisy detectors have similar performance compared to dual detectors. Hence, the case studied in our work performs the same or better for longer distances than dual detectors case because the latter suffers from lower efficiency values at these distances. For shorter implementation layouts, we derive similar results with \cite{Yuan} as demonstrated in the main text via simulations.

Through this research contribution, we incorporate the afterpulse phenomenon into the equations presented in \cite{ma2005security} succeeding to maintain the same definition for each quantity. These theoretical formulas open the door towards the study of the effect of the afterpulse phenomenon on the numerous deployed decoy-state QKD protocols. However, we mainly focus on the case of two decoy states and more specifically the weak+vacuum decoy-state where our results match the results of Fan-Yuan et al. in \cite{Yuan}, enhancing the credibility of both works. 

\section{Theoretical Results}
\label{sec:theoretical_results}

The basic equations for the decoy-state QKD protocols were introduced by Ma and Lo in \cite{ma2005security}. In this paper, we adopt the model used in \cite{ma2005security} and we build upon these equations to provide the modified ones for a step-by-step explanation of our way of thinking. For this purpose, the reader who wants to fully understand the meaning of each quantity presented below is encouraged to read the first two sections of paper \cite{ma2005security}. However, we will present the two kind of equations, the previous ones and our modified version, with this exact order in order for this paper to be as self-sustaining as possible.

\subsection{Model of source}
\label{sec:sources}

For the receiver, we have considered the general case where Bob uses $m$ detectors for the signal reception. In most cases, $m$ equals to $2$ \cite{two_detectors} and less often equals to $4$ \cite{four_detectors}. The reason behind the selection of this general case with $m$ detectors lies in the fact that different individual Single-Photon Avalanche Diode (SPAD) detectors- even if identical in type, make, brand, etc.- behave practically in a completely different way \cite{afterpulse}.

The multiple detector case is equivalent with the case of one common afterpulse probability featuring the weight of each detector:

\begin{equation} \label{eq:afterpulse_sum}
\centering
p_{\rm AP}=\sum^N_{m=1}{\dfrac{1}{N}(1+r_m)p_{{\rm AP},m}}
\end{equation}
where $p_{{\rm AP},m}$ is the afterpulse probability of the $m$ detector and $r_m$ is the bias of the $m$ detector meaning the percent change from the ideal case where every detector has the same probability to detect ($r_m=0$). These bias values may emerge from the misalignment of the setup which favours one or more detectors or from the Alice's setup which may create non-ideal qubits.

The definition implies that the following condition must hold:

\begin{equation} \label{eq:bias_sum}
\centering
\begin{split}
\sum^N_{m=1}{r_m}&=0 \\
-1 \leq r_m &\leq N-1
\end{split}
\end{equation} 
where the equalities on the latter formula for the detection $m$ signify the case that all distributing photons enter the $m$ detector ($r_m=N-1$) and the case that no distributing photon enters the $m$ detector ($r_m=-1$) respectively. The first of Eqs. \ref{eq:bias_sum} implies that when the $k^{\rm th}$ detector satisfies the $r_k=N-1$, then every other detector is forced to have $r_{m\neq k}=-1$.

\subsection{Modified representation of the general model of decoy-state QKD protocols}
\label{sec:modification}

Through our framework building, we have preserved the sources and the channel models used in \cite{ma2005security}. Hence, we present only the modified quantities connected with the detectors which are the yield $Y_i$ and the Quantum Bit Error Rate $e_i$.

\begin{itemize}
	\item Yield (\(Y_i\)): the probability that Bob detects an event conditioning an i-photon state was sent. Note that \(Y_0\) is the noise contribution \cite{ma2005security}.
	
	Considering that the main noise source is originated from the dark counts of the detector, which is very common practice, we obtain \cite{Diamanti}:
	
	\begin{center} 
		\begin{equation} \label{eqn: Y0}
		\begin{split}
		Y_0 &=p_{\rm DC} \\
		Y_i &\cong Y_0+\eta_i
		\end{split}		
		\end{equation}
	\end{center}
	where \(p_{\rm DC}\) is analogous to the number of detectors used in our setup and \(\eta_i\) is the transmittance of an i-photon state. It should be mentioned that the photons in the same state have been considered independent ($\eta_i=1-(1-\eta)^i$), while the $\eta$ includes the overall transmission and detection efficiency \cite{ma2005security}. 
	
	From the definition of the afterpulse phenomenon, we can assume that some of the detected background counts are originating from dark count events. Hence, we can modify Eq. \ref{eqn: Y0} as:
	
	\begin{center}
		\begin{equation} \label{eqn: changed Y0}
		\begin{split}
		Y_0 & = (1+p_{\rm AP})p_{\rm DC} \\
		Y_i & \cong Y_0+\eta_i(1+p_{\rm AP})
		\end{split}
		\end{equation}
	\end{center}
	where \(p_{\rm AP}\) is the already defined afterpulse probability and is provided from Eq. \ref{eq:afterpulse_sum}.
	
	\item Quantum Bit Error Rate (\(e_i\)): The i-photon state error rate is given by \cite{ma2005security}:
	
	\begin{center}
		\begin{equation} \label{eqn: old QBER}
		e_i=\frac{e_0Y_0+e_{\rm detector}\eta_i}{Y_i}
		\end{equation}
	\end{center}
	where \(e_{\rm detector}\) is the baseline system error rate \cite{Diamanti} and it describes the probability of a photon to be erroneous detected. Thus, \(e_0\) is the error rate of the background which is assumed to be random \cite{ma2005security}. Hence \(e_0=\frac{1}{2}\).
	
	When a pulse carrying single photons reaches the Bob station which hosts the SPAD units, afterpulsing may trigger one more detection event -whether or not the previous pulse was detected correctly. Furthermore, the detection triggered by afterpulsing may count as correct or wrong with equal probability. This probability is expressed by \(e_0 p_{\rm AP} \eta_i\). As a result we obtain:
	
	\begin{center}
		\begin{equation} \label{eqn: changed QBER}
		e_i=\frac{e_0Y_0+(e'_{\rm detector}+e_0p_{\rm AP})\eta_i}{Y_i}
		\end{equation}
	\end{center}
	In this case, the \(e'_{\rm detector}\) does not include the afterpulse phenomenon as in Eq. \ref{eqn: old QBER}; since we intend to isolate the afterpulsing effect; we aim to a breakdown of the basic equations. Following this path, \(e'_{\rm detector}\) can be used in our model as the quantity describing the baseline system error rate without considering the afterpulsing.
\end{itemize}

The modified formulas \ref{eqn: changed Y0} and \ref{eqn: changed QBER} describe the basic quantities for every metric of the decoy system and they can now be adopted to obtain the efficiency of the system via the total QBER and the SKR values.

Hence, the total gain, meaning the probability of having a detection, can be found:

\begin{center}
	\begin{equation} \label{eqn: new Gain}
	Q_\mu =Y_0+(1-e^{-\eta \mu})(1+p_{\rm AP})		
	\end{equation}
\end{center}
where the term \((1-e^{-\eta \mu})\) expresses the detection probability due to external photons coming from a signal state.

We are now in position to express the total QBER as:

\begin{center}
	\begin{equation} \label{eqn: new QBER}
	E_{\mu}=\frac{1}{Q_\mu}[e_0Y_0+(e'_{\rm detector}+e_0p_{\rm AP})(1-e^{-\eta \mu})]		
	\end{equation}
\end{center}

The lower bound of the Secure Key Rate is given by \cite{ma2005security}:
\begin{equation}\label{eqn:Key rate decoy}
R\geq q\{-f(E_\mu)Q_\mu H_2(E_\mu)+Q_1[1-H_2(e_1)]\}
\end{equation}
where \(f(x)\) is the bi-direction error correction efficiency as a function of the error rate and \(H_2(x)\) is the binary Shannon information function.

Equations \ref{eqn: changed Y0} and \ref{eqn: changed QBER}-\ref{eqn: new QBER} are the desired theoretical outcomes after the break down of the basic decoy-state QKD protocol. Their importance will be comprehensively discussed in the following sections.

We are able to replace our results into Eq. \ref{eqn:Key rate decoy} to estimate the lower bound of the SKR performance.

\subsection{Baseline system error rate dependence on afterpulsing} \label{sec: opt. mu}
In this section, we will derive theoretical formulas for estimating the protocol’s efficiency and for describing the dependence of the baseline system error rate from the afterpulse probability. We consider the typical case where the background rate is low (\(Y_0\ll \eta\)) and the transmittance is small \(\eta\ll 1\) as was chosen in \cite{ma2005security}.

Using our equations addresing the presence of afterpulse, the SKR can now be approximated:
\begin{center}
	\begin{equation} \label{eqn: new Key Rate}
	\begin{split}
	R\cong & -\eta \mu(1+p_{\rm AP})f(e_{\rm detector})H_2(e_{\rm detector}) \\
	&+\eta \mu e^{-\mu}(1+p_{\rm AP})[1-H_2(e_{\rm detector})]
	\end{split}			
	\end{equation}
\end{center}
which is a better approximation than the one given in \cite{ma2005security} even though the difference is very small as usually $p_{\rm AP}\ll 1$.

In the process of acquiring Eq. \ref{eqn: new Key Rate}, it was derived that the following must hold:

\begin{center}
	\begin{equation} \label{eqn: e'}			  			  	
	e_{\rm detector}=\dfrac{e'_{\rm detector}+e_0p_{\rm AP}}{1+p_{\rm AP}}	
	\end{equation}
\end{center}

Hence, a mathematical expression describing the way that afterpulsing affects the baseline system error rate was found. For \(p_{\rm AP}=0\), Eq. \ref{eqn: e'} leads to the ideal case as expected. Its correctness will be compared against results of \cite{Yuan} in Subsec. \ref{sec:implementation} through numerical simulations. Although Eq. \ref{eqn: e'} was derived through the use of the decoy-state QKD protocols model, there is no indication that it should hold only for these protocols because it connects quantities of the setup irrelevant to QKD protocols.

Furthermore, from Eq. \ref{eqn: new Key Rate}, we can find that the optimal $\mu$ fulfils:
\begin{equation} \label{eqn: new optimal mu}
(1-\mu){\rm exp}(-\mu)=\dfrac{f(e_{\rm detector})H_2(e_{\rm detector})}{1-H_2(e_{\rm detector})}
\end{equation}
which is the same as in \cite{ma2005security} but with $e_{\rm detector}$ given now by Eq. \ref{eqn: e'} so as expected the baseline system error rate, \(e_{\rm detector}\), is very important as it affects the secure key rate as shown in Eq. \ref{eqn: new Key Rate} not only directly but also indirectly through quantity \(\mu\). As a result the change, due to afterpulsing, of quantity \(e_{\rm detector}\) needs to be studied separately. 

At this point, we need to mention that if dark noise is considered sufficiently low, the QBER performance can be approximately calculated:

\begin{equation}\label{eqn: Visibility}
QBER\simeq\frac{1-V}{2}
\end{equation}	
where \(V\) expresses the interference visibility value and is a measure of the efficiency of Alice's and Bob's interferometers; representing the efficiency of separating the input pulses.

By adopting the same logical assumptions mentioned in the beginning of this subsection and using Eq. \ref{eqn: Visibility}, we can find that:
\begin{equation}\label{eqn: detector and Visibility}
V\simeq 1-2\Big[\frac{e'_{\rm detector}+e_0 p_{\rm AP}}{1+p_{\rm AP}}\Big]
\end{equation}

We confirm, as expected from the definition, that the afterpulse phenomenon affects the interference visibility and we managed to establish the connection between these quantities.

\section{Discussion and simulations}
\label{sec:discussion}
\subsection{Baseline system error rate sensitivity to afterpulsing}
\label{sec:baseline_sensitivity}

As we aforementioned the baseline system error rate (\(e_{\rm detector}\)) depends on the afterpulse probability according to Eq. \ref{eqn: e'}. We study on the change:

\begin{center}
	\begin{equation} \label{eqn: e change}
	\begin{split}
	e_{\rm detector\:change}&=\dfrac{e_{\rm detector}-e'_{\rm detector}}{e'_{\rm detector}} \\
	&=\dfrac{p_{\rm AP}}{1+p_{\rm AP}}\left( \dfrac{e_0}{e'_{\rm detector}}-1 \right)
	\end{split}		
	\end{equation}
\end{center}

Although Eq. \ref{eq:afterpulse_sum} implicates the use of more than one detectors, we will adopt the ideal case for delivering our simulations. The ideal case is where $r_m=0$ for every detector and $p_{{\rm AP},m}$ is identical for each unit. This assumption was considered since we are focusing on the afterpulsing repercussions.

Typical values of quantities \(e_0\) and \(e'_{\rm detector}\) are \(1/2\) and less than \(5\% \) respectively \cite{Lucamarini,Visibility1,Visibility2}. As a result, the afterpulse phenomenon affects significantly the baseline system error rate. Moreover, its impact becomes stronger for implementations setups featuring lower error detector performance. For \(e'_{\rm detector}=2\%\), the percent change of the total baseline system error rate as a function of afterpulsing is the following:

\begin{center}
	\begin{equation} 
	e_{\rm detector\: change} =24\dfrac{p_{\rm AP}}{1+p_{\rm AP}}	
	\end{equation}
\end{center}
where for a reasonable value of $p_{\rm AP}$ \cite{afterpulse,link2}, \(p_{\rm AP}=0.8\%\), we obtain a percent change equal to \(19\%\).

For the following simulations, we have considered similar values to \cite{Yuan} for the implementation parameters in order to make direct comparisons. More specifically, we have considered two identical detectors with total dark count probability equal to $6\times 10^{-7}$. Bob's detector efficiency is assumed to be $10\%$ and we have also considered lossless Bob's implementation setup. The optical losses of the fibers involved into the simulations are $0.21$ dB/km which is the typical attenuation parameter for the standard telecom fibers at $1550$ nm. Furthermore, the efficiency of the error correction algorithm is $1.16$.

\begin{figure}[h!]
	\centering
	\includegraphics[width=\linewidth, height=0.9\linewidth]{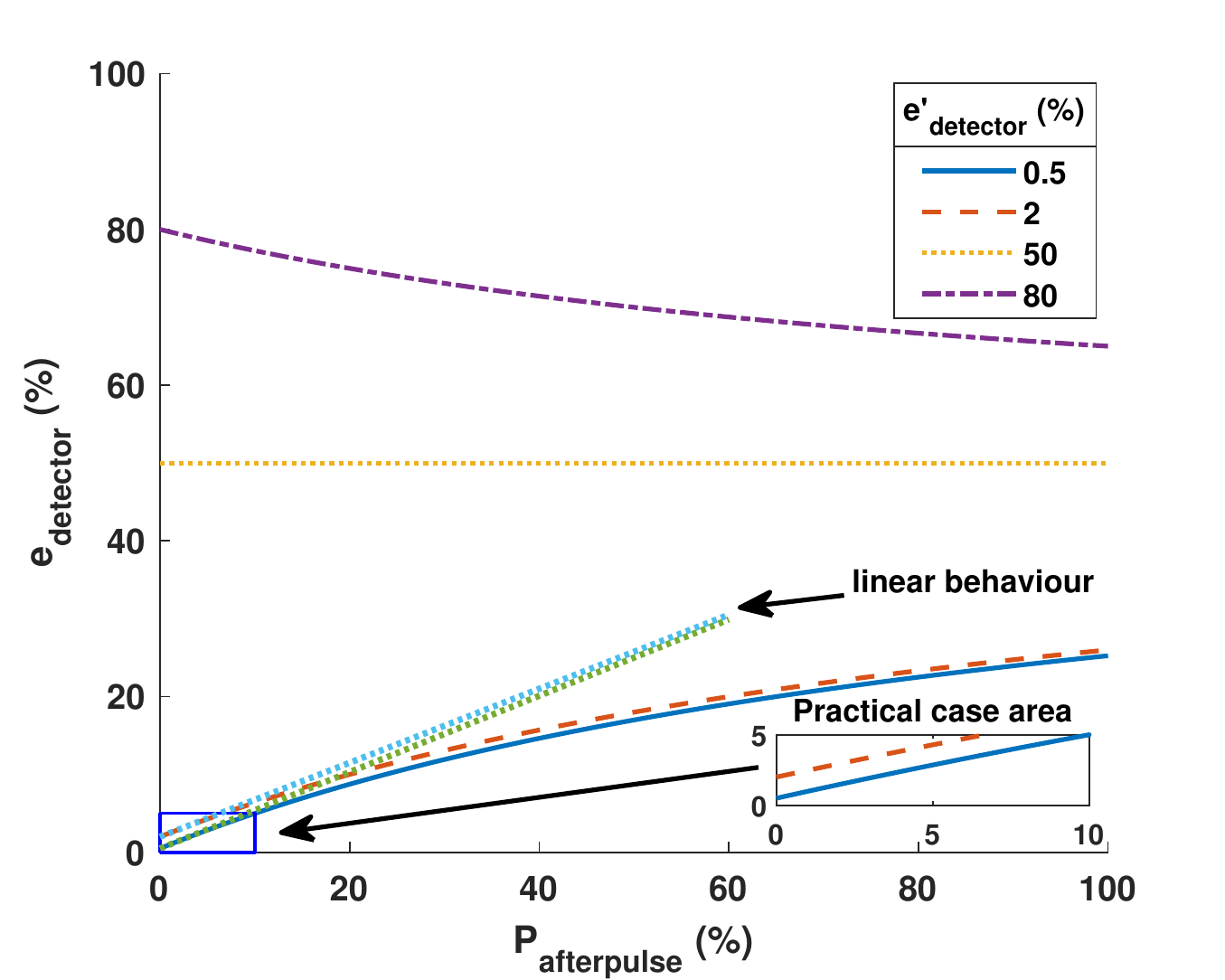} 
	\caption{Baseline system error rate related to afterpulse probability.}
	\label{fig: e_det sub1}
\end{figure}

Figure \ref{fig: e_det sub1} illustrates the baseline system error rate as a function of $p_{\rm AP}$ under different implementation parameters. Although, as we have mentioned, values of $e'_{\rm detector}$ greater than $5\%$ rarely hold in realistic deployments, we have presented them for the completeness of our study and to facilitate the discussion on some interesting perspectives of our theoretical study. When $e'_{\rm detector}>e_0$, afterpulsing seems to improve the baseline system error rate; this can be easily explained as we have incorporated a quantity (i.e. the afterpulsing) which leads to $50\%$ error rate in a quantity ($e'_{\rm detector}$) with greater error rate. It is not the first time in bibliography where unexpected features of the QKD protocols improve their efficiency \cite{Something_out_of_nothing}. Moreover, in the same figure there is a linear behaviour of the baseline system error rate (\(e_{\rm detector}\)) for varying afterpulsing within a range of afterpulse probability values (approximately less than $10\%$) and reaches the limit $50\%$ for large values of the afterpulse probability. This linear behaviour can be easily derived from Eq. \ref{eqn: e'} by taking into account that $p_{\rm AP}\ll1$. Although the relation $e'_{\rm detector}\gg e_0 p_{\rm AP}$ usually holds, it is important to preserve the second term of the numerator because due to this term $e_{\rm detector}$ is strictly increasing with the afterpulse probability, $p_{\rm AP}$, when \(e_0>e'_{\rm detector}\), which is generally true. If this term was ignored, the baseline system error rate ($e_{\rm detector}$) would be also decreasing in this case losing that way the physical meaning behind the afterpulsing parameter, i.e. the afterpulse phenomenon increases the baseline system error rate. Overall, it is reasonable that an analogous dependence behaviour between visibility and afterpulsing exists.

\begin{figure}[h!]
	\centering
	\includegraphics[width=\linewidth, height=1.8\linewidth]{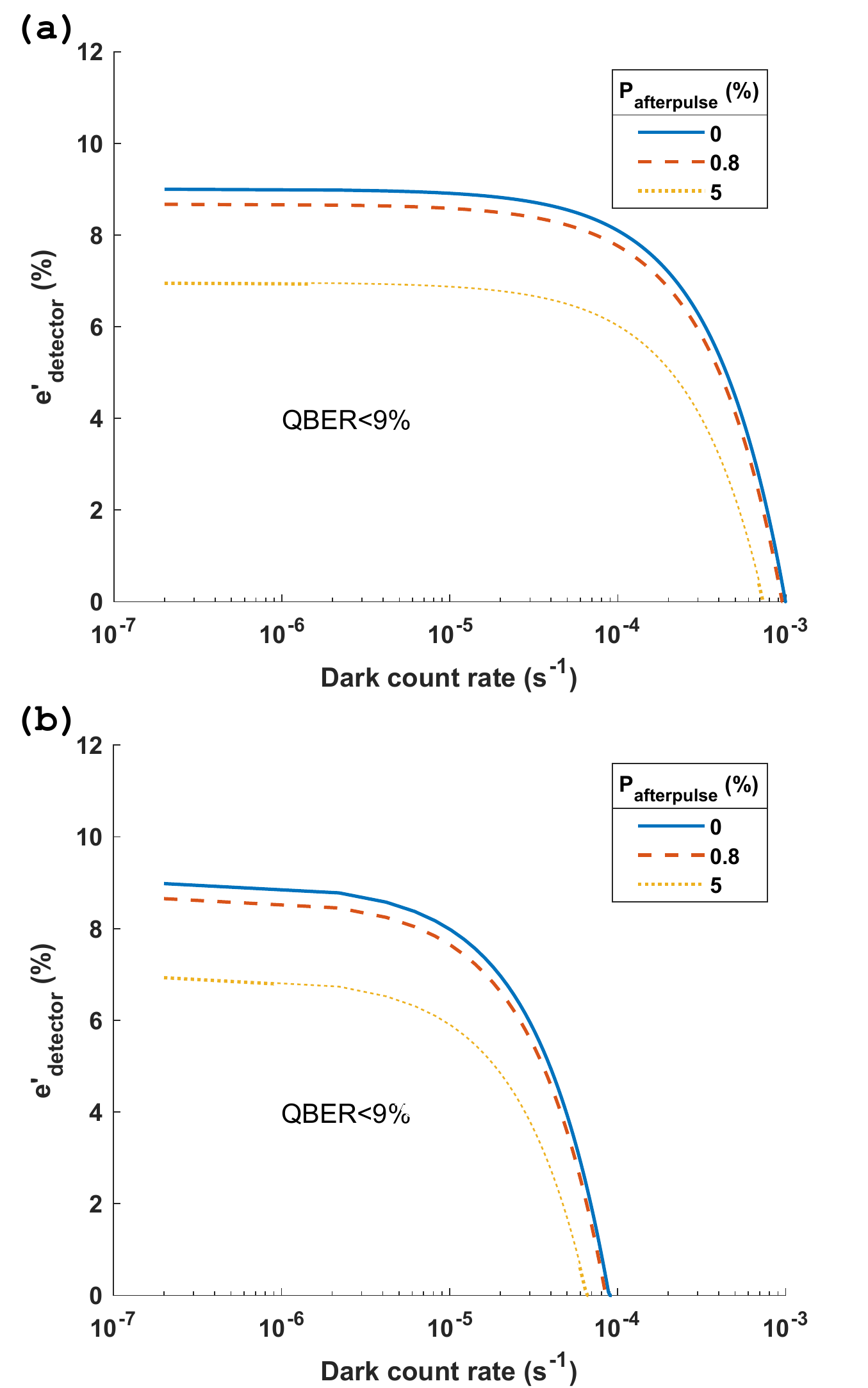}
	\caption{Relationship between the setup quantities ($e'_{\rm detector}$, dark count rate, $p_{\rm AP}$) for maintaining $QBER=9\%$ (area below graph corresponds to $QBER<9\%$). Transmission loss is $10.5$ dB and $21$ dB for subfigure (a) and subfigure (b) respectively.}
	\label{fig:QBER_e}
\end{figure}

Our mathematical model permits to precisely describe the protocols efficiency dependence on the afterpulsing and thus to establish the way that the setup's parameters affect the QBER. Then a tradeoff-like relation between the setup's parameters is emerged; the "tradeoff-like" term is used because there is a tradeoff relation if and only if the QBER has to remain the same when the parameters change, this is something that rarely applies. Figure \ref{fig:QBER_e} provides more details on the aforementioned argument. It illustrates the relationship between the setup quantities, i.e. the $e'_{\rm detector}$, the dark count rate and the $p_{\rm AP}$, for different values of transmission loss for which the QBER of the setup is equal to $9\%$ (lower than $9\%$ in the area under the graphs). This QBER threshold was chosen as it assesses the upper limit under which the frequently used CASCADE error correction algorithm works \cite{eraerds2010quantum}. Comparing the results depicted in the two subfigures of Fig. \ref{fig:QBER_e}, it is evident that the increase of the transmission loss reduces the maximum possible dark count rate that can be used for maintaining the desired QBER levels; thus there is a threshold for the dark count rate depending on the transmission loss above which whatever values the afterpulse probability and the $e'_{\rm detector}$ have, the desired QBER is not achieved. Moreover, Fig. \ref{fig:QBER_e} divulges the tradeoff-like relation among the quantities which form the total baseline system error rate- i.e. the afterpulse probability, the dark count rate and the $e'_{\rm detector}$- for maintaining the same QBER value. More precisely, keeping constant the value of one of these three quantities, the possible values of the other two quantities are tracked. This has immediate practical repercussion as one can accomplish the same QBER value by reducing the baseline system error rate (ameliorating the alignment of the setup) or the dark count rate, when only low cost detectors are available or when the SSPDs operate on high frequencies while the readout circuit is at room temperature where non zero afterpulse-like phenomenon is created \cite{afterpulse_like}. 

\subsection{Implementation of our results on the Weak+Vacuum QKD protocol}
\label{sec:implementation}

We have chosen the weak+vacuum QKD protocol since it is the most widely used implementation in the decoy-state protocols category \cite{implementation_1,implementation_2}. This protocol uses two decoy states where the first one is the vacuum state and the second one is the weak decoy state which its intensity is lower than the signal's one.

For our simulations we adopt the following assumption presented in \cite{ma2008quantum} which can be used to every decoy-state QKD protocol:

\begin{centering}
	\begin{equation} \label{eqn:error_yield}
	\begin{split}
	Y_i(\rm decoy) & =Y_i(\rm signal) \\
	e_i(\rm decoy) & =e_i(\rm signal)
	\end{split}
	\end{equation}
\end{centering}

From this assumption, we find:
\begin{center}
	\begin{equation} \label{eqn: new Gain decoy}
	Q_{\nu_1} =Y_0+(1-e^{-\eta \nu_1})(1+p_{\rm AP})		
	\end{equation}
\end{center}
where the symbol \(\nu_1\) expresses the power of the weak decoy signal and the term \((1-e^{-\eta \nu_1})\) expresses the detection probability due to external photons associated with the weak decoy state.

Then we have:
\begin{center}
	\begin{equation} \label{eqn: new QBER decoy}
	E_{\nu_1}=\frac{1}{Q_{\nu_1}}[e_0Y_0+(e_{\rm detector}+e_0p_{\rm AP})(1-e^{-\eta \nu_1})]		
	\end{equation}
\end{center}

We proceed then via simulations for visualizing the connection of the efficiency of the protocol and the afterpulse phenomenon. 

In this section, the connection between the SKR and the afterpulse probability for different transmission distances and values of baseline system error rate is aimed to be pictured. As a result, maximizing the SKR of a QKD setup demands the optimization of all other quantities involved. Those quantities are the intensities of the signal and decoy pulses. In \cite{ma2005security}, the optimized powers for the decoy pulses are found; the optimized weak pulses demonstrate an average photon number per pulse, $\nu$, equals to $0.038,\: 0.05,\: 0.12$ for $0,\: 5,\: 21$ dB transmission loss respectively whereas the average photon number per pulse of the vacuum decoy pulses ($q$) is proved to be optimized when the pulses are ideal vacuum, $q=0$. The optimized value of the average photon number for the signal state ($\mu$) depends on the afterpulse probability, on the baseline system error rate and on the transmission distance, it is different for every combination and it has been found through simulations. Hence, each figure has the best possible combination of the power of each of the three states of the weak+vacuum protocol for the given detector's quantities and the transmission distance in order to study only the effect of the afterpulse probability.

\begin{centering}
	\begin{figure}[h!]
		\includegraphics[width=\linewidth, height=1.8\linewidth]{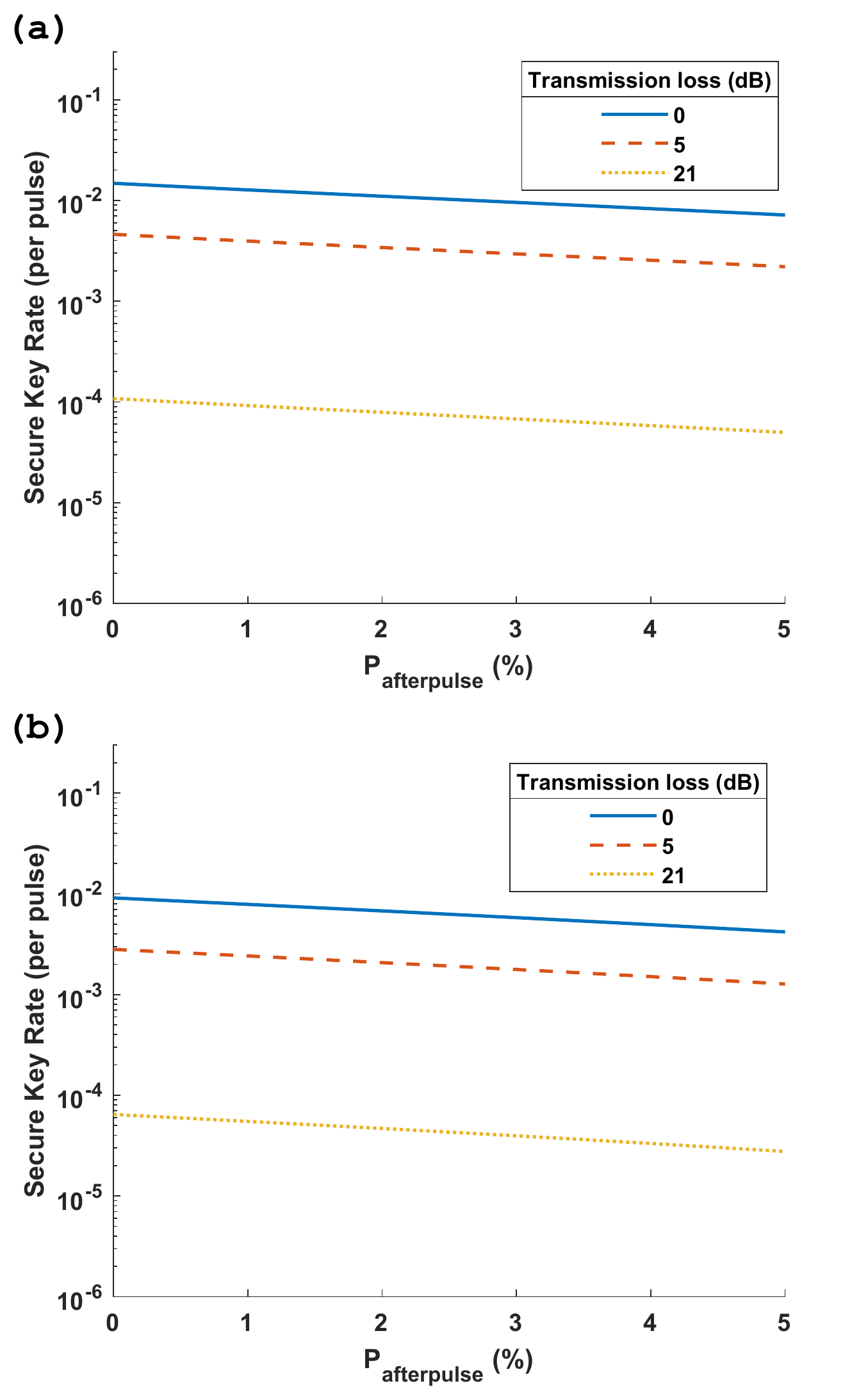}
		\caption{Secure Key Rate per pulse in relevance with the afterpulse phenomenon in a logarithmic scale for different base line error rate in each subfigure. Subfigure (a) is for $e'_{\rm detector}=0.5\%$ and subfigure (b) is for $e'_{\rm detector}=2\%$.}
		\label{fig:Rmax parts}
	\end{figure} 
\end{centering}

Figure \ref{fig:Rmax parts} represents our results. In Fig. \ref{fig:Rmax parts}(a) the same baseline system error rate as in \cite{Yuan} was used; comparing the cases of $0$ dB and $5$ dB transmission losses with the corresponding figure in \cite{Yuan}, an almost exact agreement is visible enhancing both works credibility. These two subfigures indicate as expected that the higher the baseline system error rate without the afterpulse probability ($e'_{\rm detector}$) is, the smaller the generated SKR is.

\section{Conclusions}
\label{sec:conclusions}

In this work we have broken down the various effects of the afterpulsing on every quantity involved in decoy-state QKD protocols with undefined number of different detectors. An additional bias favouring one or more detectors compromising the rest is considered, generalizing Bob's setup.

Through the process of extracting the aforementioned theoretical formulas, the dependence between the baseline system error rate and the afterpulse probability was revealed as a necessity for compatibility between the equations presented in \cite{ma2005security} and our modified ones. However, we believe that this relation also holds outside the decoy-state protocols as it describes a setup quantity and as it only relies on other setup implementation quantities and not on the protocol's parameters; it is for future experiments to ascertain this argument. Furthermore, the value of the total baseline system error rate has a major impact on the QBER of the system, revealing a tradeoff-like relation among its components such as the afterpulse probability and the dark count rate. Numerical simulations studied -using realistic values for each quantity- to trail this tradeoff-like relation where also a transmission loss dependent upper threshold for the dark count rate is revealed, above which the desired QBER cannot be accomplished. A similar dependence between the visibility of the setup and the afterpulsing is also proved to exist.

Therefore, we have demonstrated a way to study \textit{\`{a} priori} the effect of the afterpulse phenomenon on the setup's efficiency for the general case of $N$ detectors making feasible the proper choice for balancing the detectors' quality and their cost. Applying our results to the weak+vacuum decoy-state QKD protocol, the outcomes of our approach were verified by comparing the SKR performance of previous related work under the same assumptions; enhancing that way all of our outcomes, i.e. both the aforementioned breakdown of the general decoy-state QKD model and the derived mathematical expression for the baseline system error rate and the afterpulsing.

We believe that the methodology presented in our research, for the case of the afterpulse phenomenon, can be a guidance for also incorporating other future imperfections in a single unified formula.

\section*{Acknowledgments}
We would like to thank Themistoklis Mavrogordatos for useful discussions and comments.

\section*{Authors' contributions} 
C. Papapanos is the corresponding author of this research. C. Papapanos conceived and comprehended the presented idea and developed the theory as well as the analytical calculations.
D. Zavitsanos, A. Raptakis and G. Giannoulis were, also, involved supporting the literature survey and the preparation of the manuscript. All the authors have read and approved the final manuscript.

\bibliography{refer}

\end{document}